\input{aipcheck}

\documentclass[
    ,final            
  ]
  {aipproc}

\layoutstyle{8x11single}

\begin{document}

\title{Peripheral NN scattering from subtractive renormalization of chiral interactions}


\author{E. F. Batista}{
  address={Departamento de Ci\^encias Exatas e Naturais, Universidade Estadual do Sudoeste da Bahia\\ 45700-000 Itapetinga - BA, Brasil }
}

\author{S. Szpigel}{
  address={Centro de R\'adio-Astronomia e Astrof\'\i sica Mackenzie, Escola de Engenharia \\ 
  Universidade Presbiteriana Mackenzie \\
              01302-907 S\~ao Paulo - SP, Brasil }
}

\author{V. S. Tim\'oteo}{
  address={Grupo de \'Optica e Modelagem Num\'erIca - GOMNI, \\ Faculdade de Tecnologia - FT, 
  Universidade Estadual de Campinas - UNICAMP \\13484-332 Limeira - SP, Brasil}
}

\begin{abstract}
We apply five subtractions in the Lippman-Schwinger (LS) equation in order to 
perform a non-perturbative renormalization of chiral N3LO nucleon-nucleon interactions. 
Here we compute the phase shifts for the uncoupled peripheral waves at renormalization 
scales between $0.1~ \rm{fm}^{-1}$ and $1 ~ \rm{fm}^{-1}$. In this range, the results are 
scale invariant and provide an overall good agreement with the Nijmegen partial wave 
analysis up to at least $E_{\rm{lab}} = 150 ~ \rm{MeV}$, 
with a cutoff at $\Lambda = 30~\rm{fm}^{-1}$.
\end{abstract}

\maketitle


\section{Introduction}

The problem of the non-perturbative renormalization of the nuclear force in chiral effective field theory(ChEFT) has been intensively investigated by several authors \cite{bedaque1,epelbaum1a,machleidt1,epelbaum1b}, creating a deep discussion regarding the consistency of the scheme originally proposed by Weinberg \cite{weinberg1,weinberg2,weinberg3}. One of the most common procedures for the non-perturbative renormalization of the nuclear forces in the context of Weinberg's approach to ChEFT can be divided in two steps \cite{epelbaum1a}. In the first step, one has to solve a regularized Lippmann-Schwinger (LS) equation for the scattering amplitude by iterating the effective $NN$ potential truncated at a given order in the chiral expansion, which includes long-range contributions from pion exchange interactions and short-range contributions parametrized by nucleon-nucleon contact interactions. In the second step, one has to determine the strengths of the contact interactions, the so called low-energy constants (LECs), by fitting a set of low-energy scattering data. Once the LEC's are fixed at a given momentum cutoff scale, the LS equation can be solved to give the on-shell T-matrix so we can derive other observables. 

The standard approach used to regularize the ultraviolet (UV) divergences in the LS equation is to introduce a sharp or smooth momentum cutoff regularizing function \cite{epelbaum1a,machleidt1} that suppresses the contributions from the potential matrix elements for momenta larger than a given momentum cutoff scale. The multi-pion exchange interactions also contain UV divergent loop integrals which must be consistently taken care of. The $NN$ interactions can be considered properly renormalized when the predicted observables are (approximately) independent of the momentum cutoff scale within the range of validity of the effective theory  \cite{machleidt1,lepage}.

The cutoff approach yields to a very efficient regularization of the interaction provided the cutoff scale 
remains below 1 GeV. For larger cutoffs, there is no convergence since the force becomes extremely 
strong for large momenta. The subtractive renormalization applies a different philosophy for the
renormalization  procedure: the scattering equation is modified and the interaction is not cut. 
The subtracted kernel makes the scattering amplitude finite.

In the next sections we briefly present the renormalization scheme with five subtractions for the chiral N3LO 
interactions, show our results for the uncoupled peripheral waves and give our main conclusions and outlook.

\section{Uncoupled channels with five subtractions}
\label{sub}

The phase shifts for the uncoupled channels can be obtained from the reaction matrix. Since the N3LO
interactions contain terms up to ${\cal O}(q^4)$, a finite K-matrix can be obtained by making five subtractions 
at a given momentum scale $\mu$. The LS equation for the on-shell K-matrix with five subtractions is given by  
\cite{npa99,plb00,plb05,aop2010,prc11,jpg2012}

\begin{eqnarray}
K^{(5)}_{\mu}(k,k) &=& V_{\mu}^{(5)}(k,k) + \frac{2}{\pi} \mathcal{P} \int_0^\infty dq~q^2~
V_{\mu}^{(5)}(k,q)~G^{(5)} (q^2; k^2, \mu^2)~K^{(5)}_{\mu}(q,k) \; ,
\label{K5}
\end{eqnarray}
where $\mathcal{P}$ denote the principal value and the driving term $V_{\mu}^{(5)}(p,p')$ 
has to be obtained recursively:
\begin{equation}
V_\mu^{(n)} (p, p'; k) = V_\mu^{(n-1)}(p, p'; k) + \frac{2}{\pi} \int_0^\infty dq~q^2~V_\mu^{(n-1)}(p, q; k)~ \frac{(\mu^2+k^2)^{n-1}}{(k^2+q^2)^n}~ 
V_\mu^{(n)}(q, p'; k)   \;  ,
\label{Vn}
\end{equation}
and
\begin{equation}
G_\mu^{(5)} (q^2; k^2) =  \frac{{\cal R_\mu}^{(5)}(q^2;k^2)}{k^2 - q^2} \; ,
\end{equation}
is the five times subtracted Green's function which contains an energy-dependent hyperbolic 
regularizing function,
\begin{eqnarray}
{\cal R}_\mu^{(5)}(q^2;k^2) = \left(  \frac{\mu^2+k^2}{\mu^2+q^2}   \right)^5  \; ,
\end{eqnarray}
that acts in the recursive driving term $V_{\mu}^{(5)}(p,p')$. Note that this is completely different 
of applying a regularizing function directly to the potential $V_{\rm{N3LO}}(p,p')$.

Here we consider two state-of-the-art N3LO chiral potentials: the one developed by 
Entem and Machleidt (N3LO-EM) \cite{mach2003} and the one developed by 
Epelbaum, Gl\"ockle and Meissner (N3LO-EGM) \cite{epel2005}. The main difference
between these two chiral forces is the calculation of the two-pion exchange component. 
In the N3LO-EGM version, the loop integrals are regularized with the use of spectral functions 
controlled by a $\tilde\Lambda$ parameter. The spectral function regularization softens the 
two-pion-exchange contribution and improves the convergence of the chiral expansion for
the nucleon-nucleon interaction \cite{egm2004a,egm2004b}.  

Each term of the chiral expansion enters in the recursive process in the step required to renormalize it.
Hence, the LO contribution goes into the first subtraction and the NLO terms goes into the third subtraction due to the polynomial contact interactions. The N2LO two-pion exchange enters into the fourth subtraction 
and the N3LO corrections go into the fifth subtraction which accounts for the renormalization of the 
${\cal O}(q^4)$  part of the nucleon-nucleon interaction.

Once the K-matrix is obtained through the numerical solution of Eq. (\ref{K5}), the phase shifts for the uncouple peripheral waves can be computed using the well known relation
\begin{eqnarray}
{\rm tan} ~ \delta^{(5)}_\mu (k) = -k~K^{(5)}_{\mu} (k,k) \; .
\end{eqnarray}
The numerical solution of the subtracted LS equation Eq. (\ref{K5}) and of the integral equations for the 
recursive driving terms Eq. (\ref{Vn}) are carried out by discretizing the momentum space 
with 200 points up to 30 $\rm{fm}^{-1}$. 

\begin{figure}
   \includegraphics[height=.2\textheight]{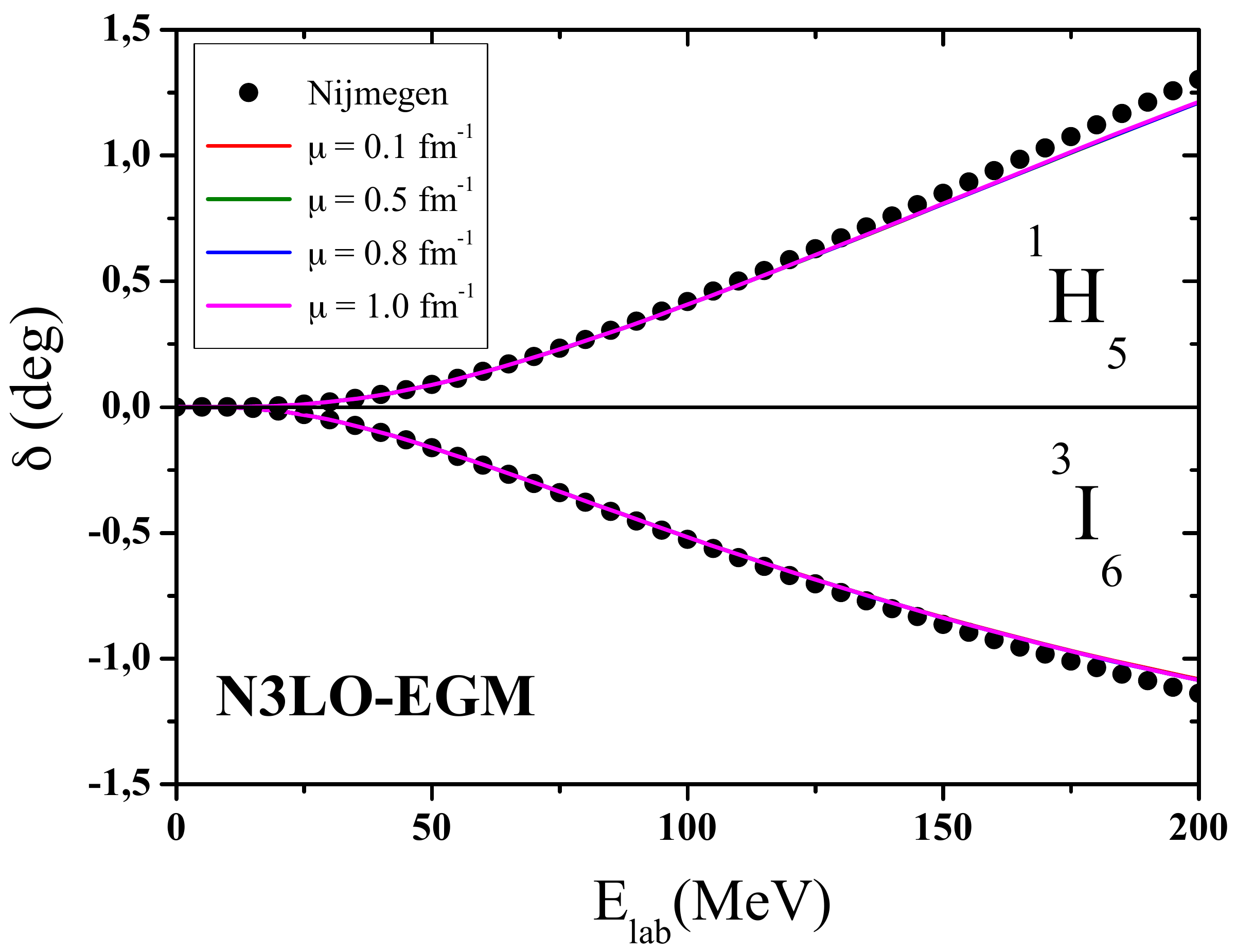}
    \hspace{1cm}
  \includegraphics[height=.2\textheight]{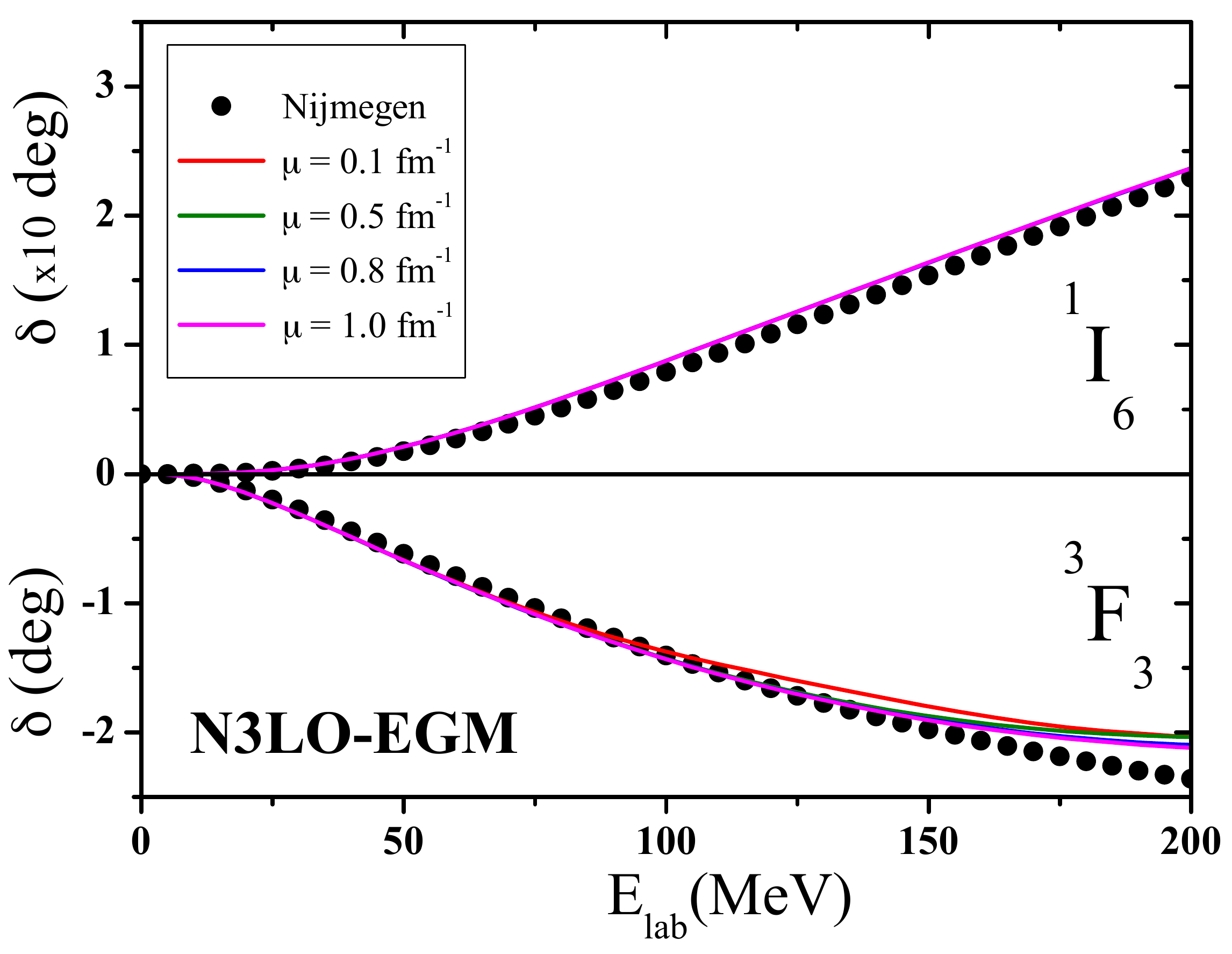} 
  \caption{Phase shifts in uncoupled peripheral waves with 
  the N3LO-EM potential (four upper panels) and N3LO-EGM (four lower panels) for subtraction 
  points below $\mu = 1~\rm{fm}^{-1}$.}
  \label{fig}
\end{figure}

In the first four panels of Figure \ref{fig} we show the phase shifts for the uncoupled peripheral 
waves up to $J=6$ 
for renormalization scales below $1~\rm{fm}^{-1}$ for the N3LO-EM potential compared to the 
Nijmegen partial wave analysis \cite{nijmegen}. Considering that we are using an extreme large cutoff, the agreement 
with the data is very good up to 200 MeV except for the F-waves which have good description only 
up to 100 MeV.

The results for the N3LO-EGM potential are displayed in the last four panels of Figure \ref{fig}, 
where we observe an improvement 
of the F-waves due to the weaker two-pion exchange contributions. For the other waves, the phase shifts 
are very similar to the ones obtained with the N3LO-EM potential.

For both potentials we can observe an approximate scale invariance in the case of the peripheral 
waves even with such large cutoff in the LS equation. This indicates that the renormalization with 
multiple subtractions is suitable for non-perturbative renormalization of chiral interactions no matter 
how large the cutoff is taken.
%
%
\begin{figure}
  \includegraphics[height=.2\textheight]{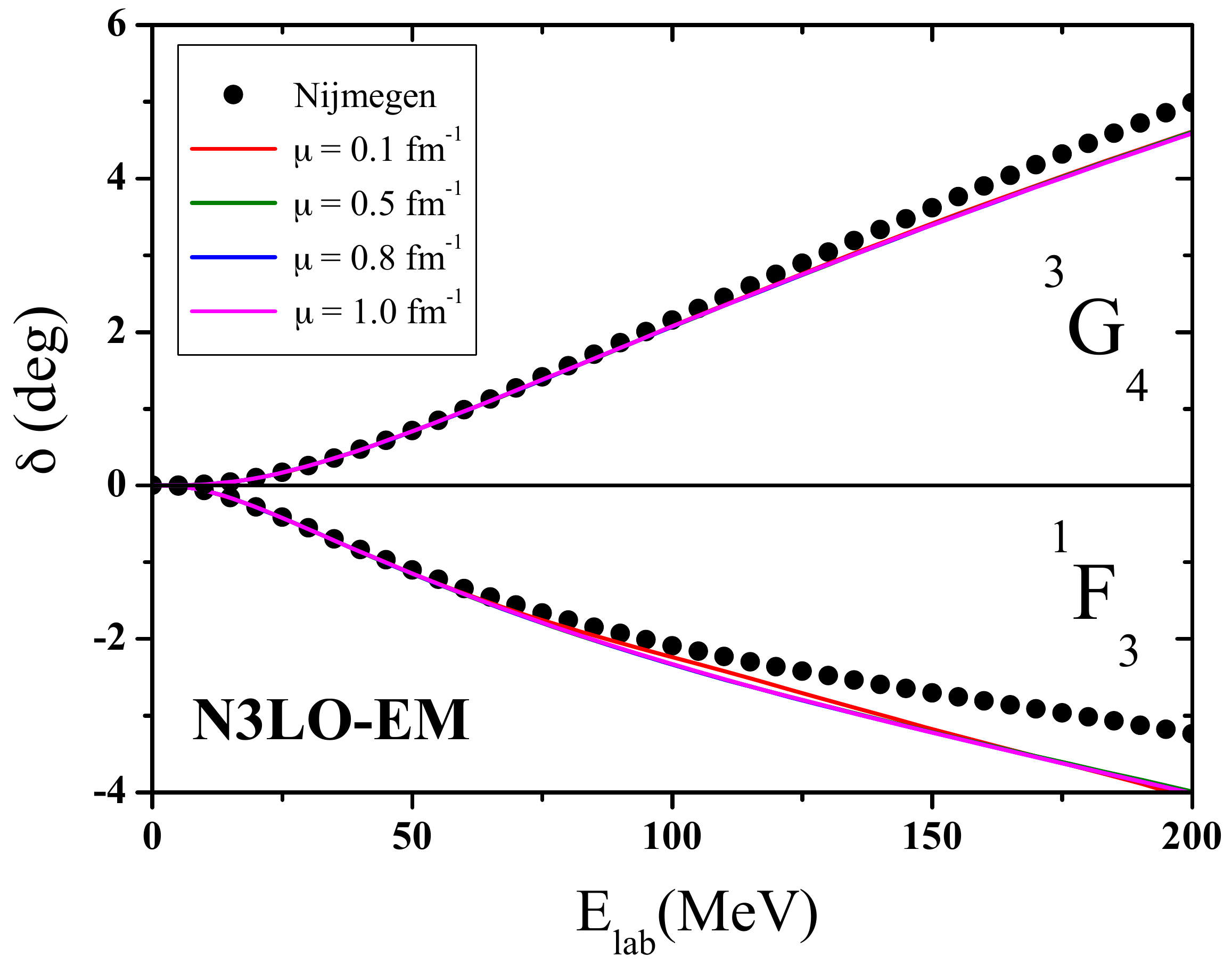}
  \hspace{1cm}
  \includegraphics[height=.2\textheight]{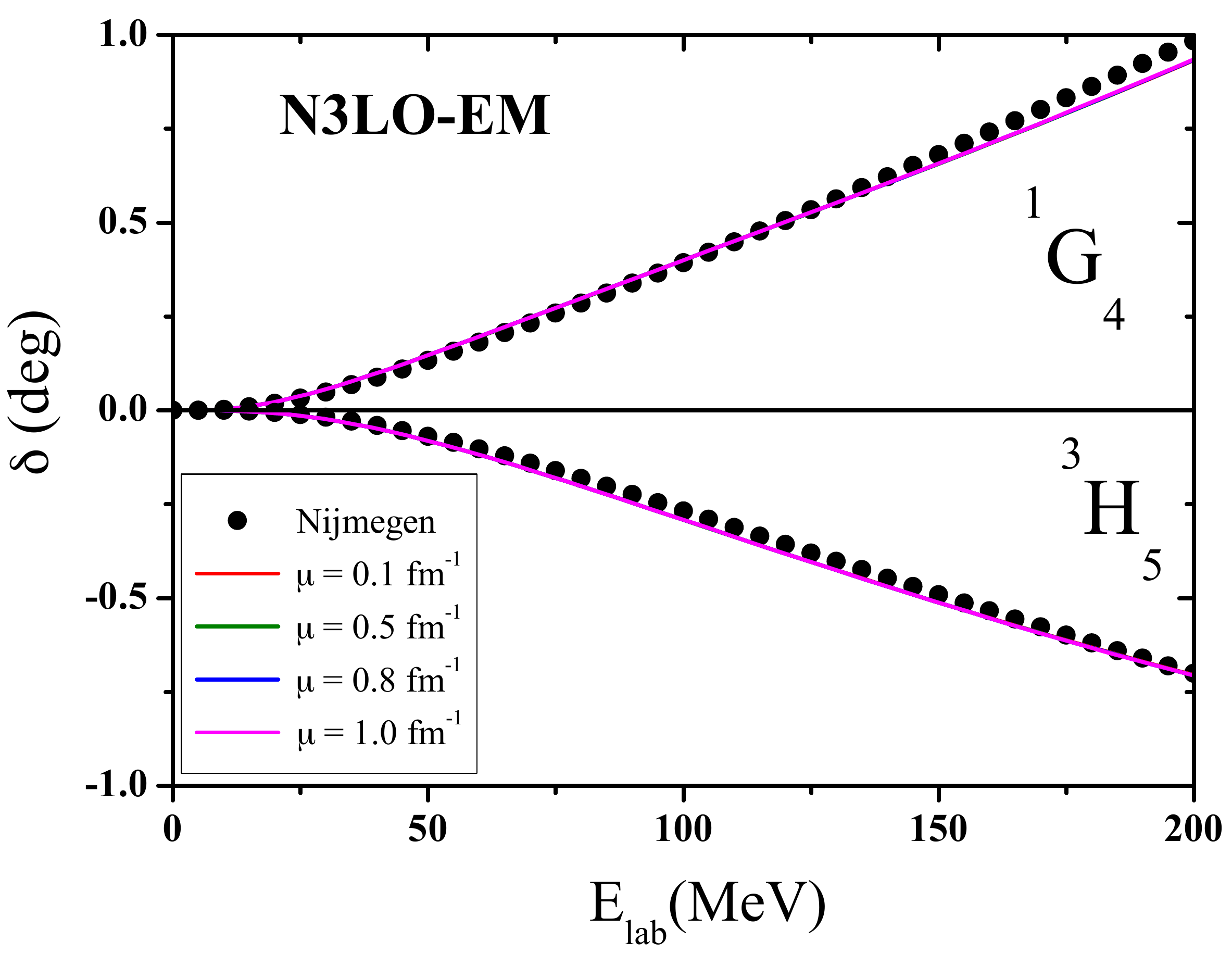} 
\end{figure}
\begin{figure}
   \includegraphics[height=.2\textheight]{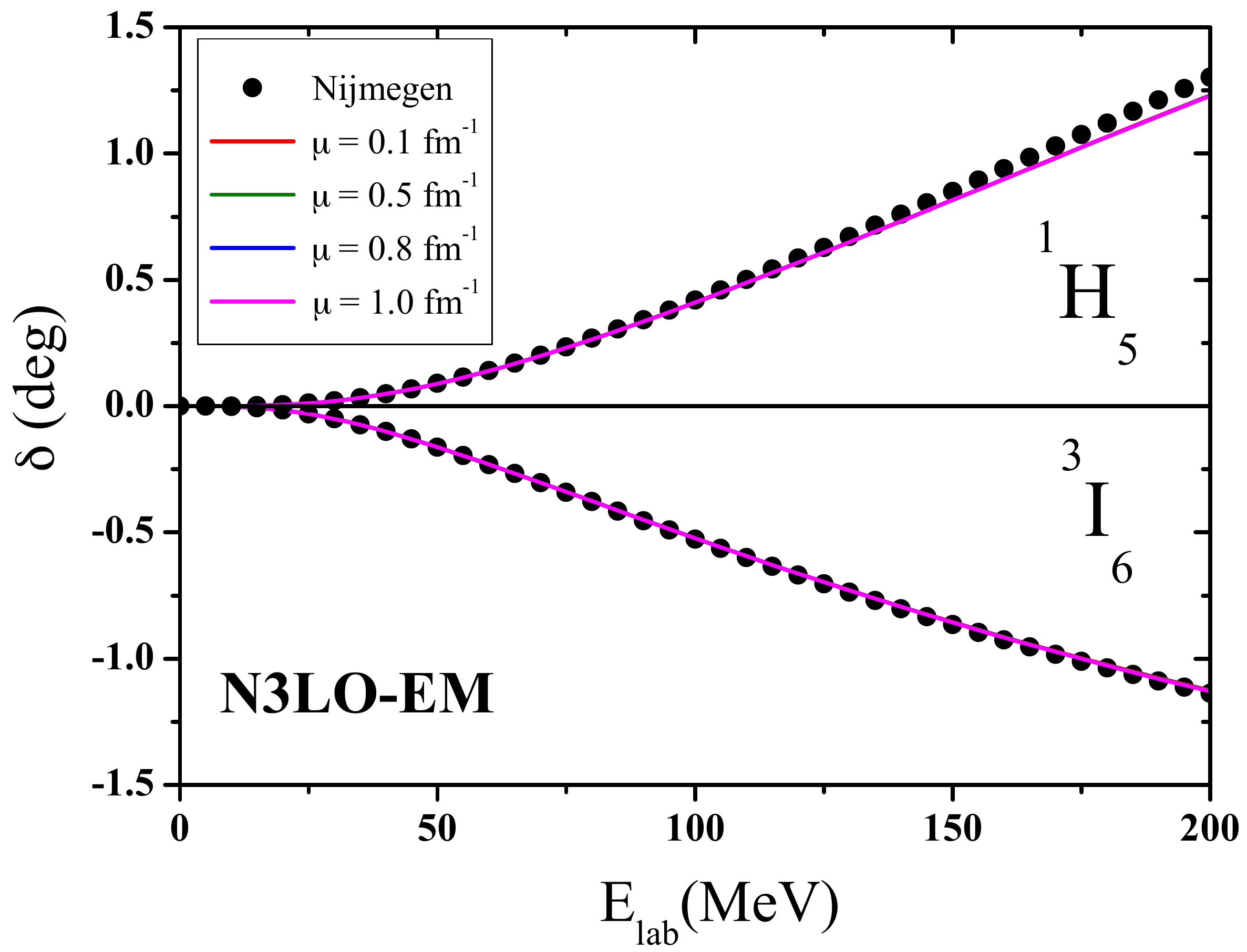}
    \hspace{1cm}
  \includegraphics[height=.2\textheight]{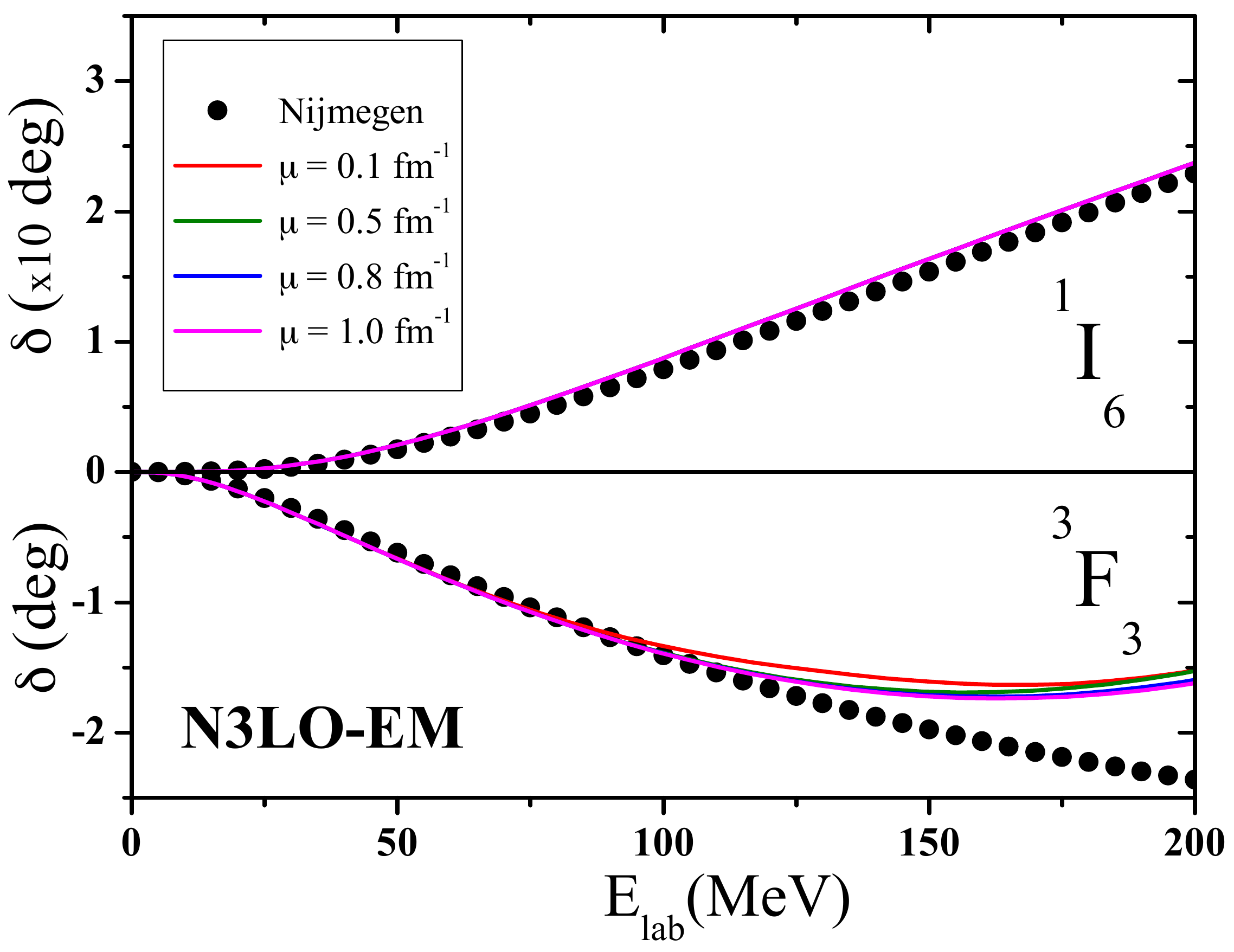} 
\end{figure}
%
%
\begin{figure}
  \includegraphics[height=.2\textheight]{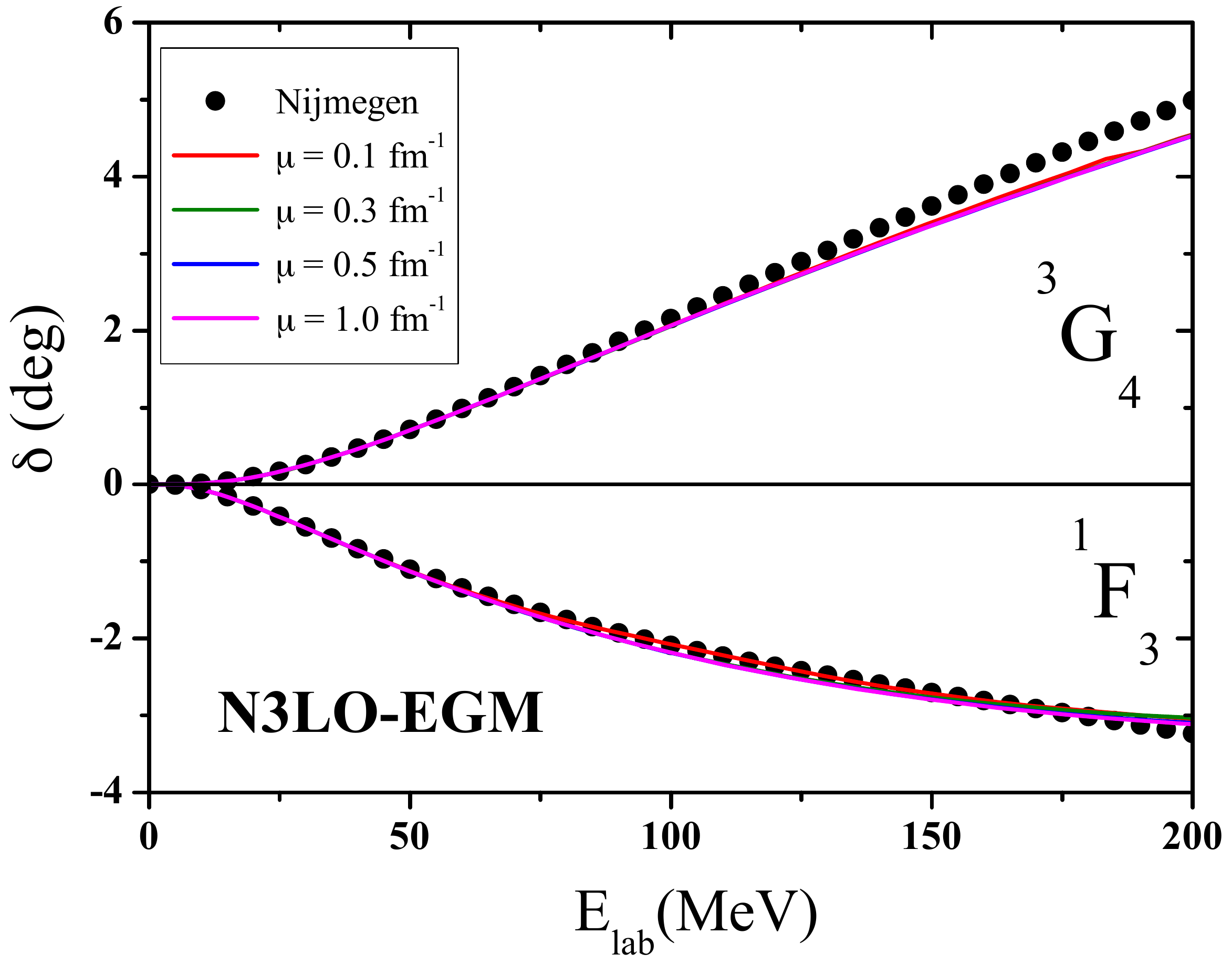}
   \hspace{1cm}
  \includegraphics[height=.2\textheight]{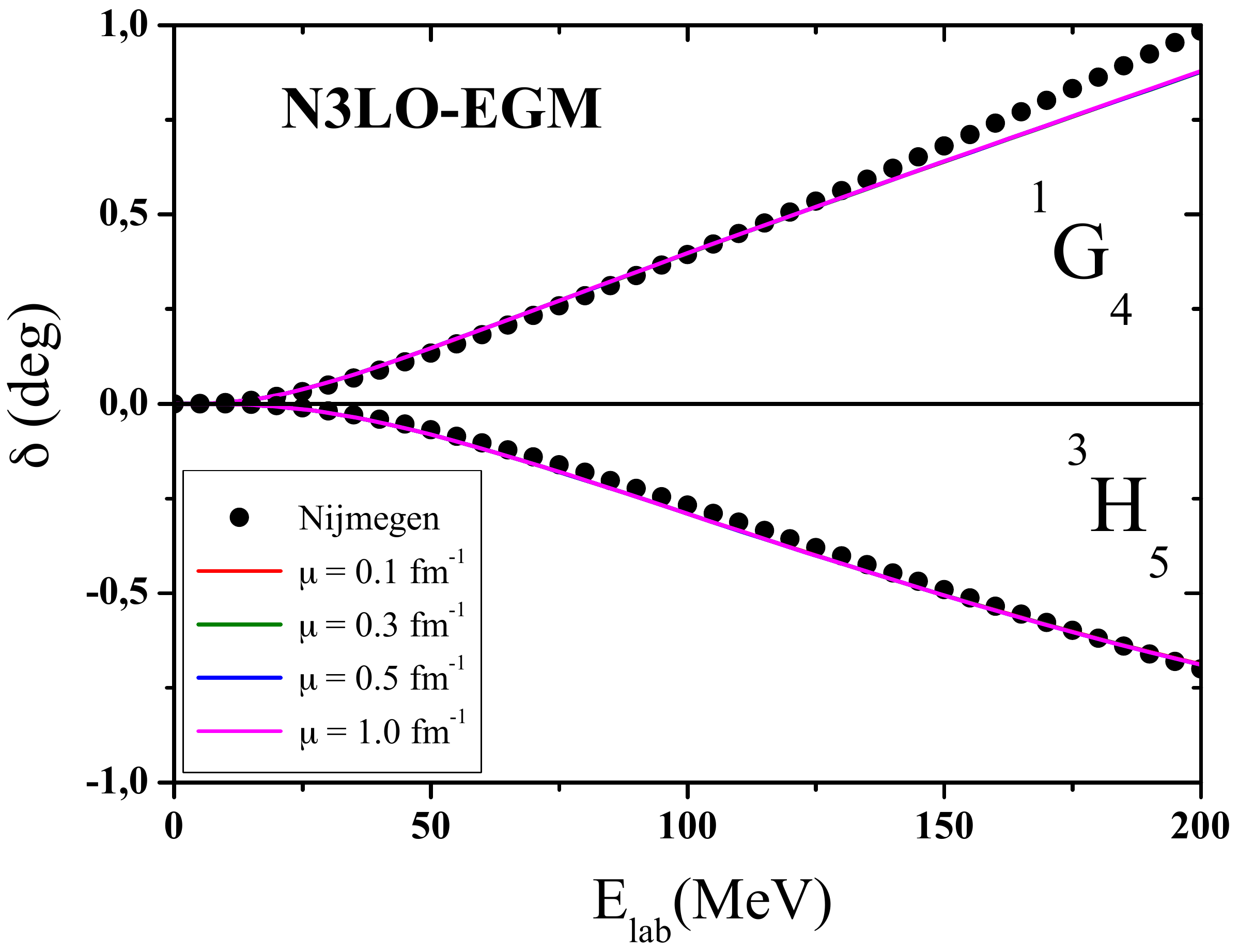} 
\end{figure}

\section{Concluding remarks}

We have implemented a non-perturbative renormalization method based on with multiple subtractions 
to compute the phase shifts for the uncoupled peripheral waves up to $J=6$. 
Our results show that the method is robust enough to describe the data with an overall good agreement
up to laboratory energies of about 200 MeV. Also, the dependence of de results on the renormalization scale is very weak below 
$\mu = 1~\rm{fm}^{-1}$ in the case of the uncoupled peripheral waves. 
If one considers that we are performing a non-perturbative renormalization of N3LO 
chiral interactions at a very large cutoff, $\Lambda = 30~\rm{fm}^{-1}$, the results indicate 
that the renormalization with multiple subtractions provides a robust technique to deal with the nuclear force.



\begin{theacknowledgments}
The authors would like to thank FAEPEX, FAPESP and CNPq for financial support. Computational
resources provided by FAPESP grant 2011/18211-2.
\end{theacknowledgments}

\bibliographystyle{aipproc}   

\end{document}